\documentstyle[11pt]{article}
  \input epsf
\newcommand{\ra}{\rightarrow}   % ->

\newcommand{\RR}{\mbox{\protect\makebox[.15em][l]{I}R}}
\newcommand{\ZZ}{\mbox{\protect\makebox[.15em][l]{\sf Z}\sf Z}}

\newcommand{\halmos}{\hfill$\Box$}
\newcommand{\rhalm}{{\bf\rfloor^{\hspace{-2.7mm}<}}}
\newcommand{\lhalm}{{\bf\lceil_{\hspace{-1.2mm}>}}}
\newcommand{\Matlab}{{\sc Matlab }}

\newcommand{\sbs}{\subseteq}
\newcommand{\sps}{\supseteq}

\newcommand{\y}{\hspace*{5mm}}
\newcommand{\hly}{\hspace*{1.63mm}}
\newcommand{\hy}{\hspace*{2.5mm}}

\newcommand{\hhy}{\hspace*{1.25mm}}
\newcommand{\nny}{\hspace*{-1.25mm}}

\begin{document}

\begin{center}
{\bf \Large  Convex Hull Calculations: \\

 a \Matlab Implementation and Correctness Proofs for  the    lrs-Algorithm } 

\vspace{.5cm}

Alexander Kova\v{c}ec, Dep. Math. Univ. Coimbra, 3001-454 Coimbra, Portugal.
 {\tt kovacec@mat.uc.pt} \\
Bernardete Ribeiro, Dep. of Informatics Engineering, Coimbra, 3030 Coimbra, Portugal.
 {\tt bribeiro@dei.uc.pt}

Dedicated to Professor G. N. de Oliveira on his sixty-fifth birthday. 

\end{center}

Abstract. This paper provides full  \Matlab-code and informal correctness proofs for 
the lexicographic reverse search algorithm for convex hull calculations. The 
implementation was tested on a 1993  486-PC for various small and some larger,  
partially highly degenerate combinatorial polytopes, one of which    (a certain
13-dimensional  24 vertex polyhedron) 
occurs naturally in the study of a well known problem posed by Professor 
Graciano de Oliveira: see end of section 1. \\

{\bf 0. Introduction } 

David Avis [A] has recently published an improvement of his 1992 reverse search 
algorithm, called lexicographic reverse search, {\it lrs} for short, for 
mediating between the VR- and the H-representation of pointed polyhedra.
[A] explains the ideas of the algorithm and and gives a 15 lines pseudo code. 
The algorithm was implemented in C-code and can be downloaded from its home
page in the vicinities of {\tt www.cs.mcgill.ca } (McGill University 
Computer Science Department). Click on people, Avis, lrs home page. 

To compile this or competing codes (like PORTA), users  have  nowadays 
to acquire one of the more modern `platforms', that, by cramming disks full with
superfluous and (ob)noxious toys serve mainly - and explicitly, see (the article 
containing) [M] - the goal to make people throw away their old, beloved, and still 
useful PCs, they finally  got to dominate. Disdaining firms that deride their customers  
with evermore rapid `updating' cycles,
authors  recently implemented  {\it lrs} in a version that 
would run on an           `old' 1993 486-PC with  equally `outdated' \Matlab for Windows 3.1. 
\Matlab  users have now an algorithm that, different from the Quickhull implementation 
on recent \Matlab 6.1 (not running on Windows 3.1 anyway), will not indicate 
the three dimensional cube, for example,   as  simplicial.  

In this paper we describe our \Matlab implementation and its usage. 

In section 1 we describe, with examples, its use for those who  more or less 
blindly wish to copy the code and use it. 
In section 2 we assume the reader to have   
[A] at hand and read. The        specifics of our implementation 
are given quite detailed explanations that should convince of the
correctness of the 
code  given in section 4. These explanations do not cover syntactical 
details nor can they guarantee numerical error freeness for too large 
polyhedra (i.e. too many rays or vertices or facets). 
It is helpful, of course, the reader has a smattering 
knowledge  of \Matlab\nny; however, if he has not, section 3 presents the 
bare minimum necessary to understand the code.  The proofs and 
examples should be valuable in case there remain any bugs, for illuminating 
Avis' paper from further perspectives, for extensions and refinements of the 
code itself, and for implementing {\it lrs} 
in still other languages.

Disclaimer: users that wish to do series of experiments with really large 
and perhaps `pathological' polyhedra, 
as may be the case in the optimization community, are for reasons of 
speed and numerical correctness
probably well advised to stick with Avis' C-implementation in  
exact arithmetic. In our implementation we used a simple heuristic for suppressing
`numerical noise'. In spite of the  warning [A, p194c3] (=page 194, about 3cm 
from first text row) that 
`the reverse search is extremely sensitive to numerical error', 
for reasons we do not quite   understand, the heuristic seems to work well 
for the  `reasonable' rational and not too large polyhedra about      
which the pure mathematician wishes information. In case the user senses problems, 
it might be useful to  choose tolerance settings different from ours, or to do  
different runs of {\it lrs } on the same polytope. \\

{\bf 1. Use of the \Matlab implementation of lrs}

For the explanations to follow below we recall some facts from the general 
theory of polyhedra. We use notions meticulously as defined in 
Schrijver [Sch], or Ziegler [Z].  Indeed the reader not too familiar with 
polytopes and polyhedra, may wish first of all 
to recall/transcribe  the following notions and associated notation: 
feasible  set of conditions [Sch, p4c1];  
polyhedral cone, finitely generated cone, polyhedron, [Sch, p87...89]; 
characteristic cone and lineality space of polyhedron, pointed polyhedron, 
[Sch, p100]; face and facet, [Sch, p101]; minimal face and vertex of polyhedron [Sch, p104];
edge and extreme ray of pointed polyhedron,  minimal proper face of cone [Sch, p105].

One form of the decomposition theorem for polyhedra in full generality reads as follows, 
the uniqueness claims at the end being a consequence of the discussion in [Sch, p107].

{\bf 1.1 Theorem } Let $P=\{x: Hx \leq b\}$ be a nonempty polyhedron. \\
i. Choose for each minimal face $F$ of $P$ a point $x_F\in F.$ \\
ii. Choose for each minimal proper face $F$ of $\mbox{char.cone} P$, an arbitrary vector  
   $y_F\in F\setminus \mbox{lin. space} P$ \\
iii. Choose and arbitrary collection $z_1,\ldots, z_t$ of vectors generating 
    $\mbox{lin. space}\, P.$ Then
\vspace{-.2cm}
\begin{eqnarray*}
P&=&\mbox{conv.hull} \{x_F: \mbox{ $F$ minimal face of $P$ }\}
 +\mbox{cone} \{ y_F:\mbox{$F$ minimal proper face of \hly char.cone $P$} \}\\ 
 & & \mbox{}+\mbox{lin.hull}\{z_1,\ldots,z_t\}. 
\end{eqnarray*}
The sets conv.hull$\{...\},$  cone$\{...\},$ and lin.hull$\{...\}$ herein are unique. \halmos

Ideally one could hope that software dealing with polyhedra could mediate between 
these representations in full generality, indicating, given $H,b,$  full sets of
generators for the three sets referred in the uniqueness claim, and conversely. 

Avis [A] and we, in this paper,  settle  for slightly less, excluding polyhedra 
like infinite prisms, halfspaces etc: the current implementation 
of {\it lrs} works for pointed polyhedra (i.e. above $t=0$); or equivalently for polyhedra that have at least 
one vertex; see [Sch, p104c-7]. This requirement is natural since the first  preprocessing
step for {\it lrs} requires translating the inequality system $Hx\leq b$ into a 
{\it dictionary}. This is close to  the simplex tableau [Sch, p133c6] and 
necessitates   a vertex for definition. 
For pointed polyhedra, the minimal faces are precisely the vertices, the minimal 
proper faces of characteristic cones are extreme rays [Sch, p105c-0].
Modifying slightly terms introduced 
in [Z], we call the representations of a pointed polyhedron $P$ in the 
hypothesis and conclusion of the  
decomposition theorem a H-representation and a VR-representation respectively; 
here H stands for `halfspace', VR for `vertex and (extreme) rays' ([Z] uses 
the term V-representation and V there stands for `vectors'). It is now    
plain that our VR-representation is equivalent to Avis 
[A, equation p181c5].
We can now be 
precise on what {\it lrs} does: 
it allows to translate a H-representation of a pointed
polyhedron into a VR-representation (via calling {\tt [V,R]=htovr(H,b,initv)} )
and a VR-representation of such a polyhedron into a H-representation (via 
calling {\tt [H,b]=vrtoh(V,R)}).

We describe these uses  by means of two examples 

{\bf 1.2 Example}  From a  H-representation to a VR-representation. \y
On [A, p181] we find a H-representation of a $3$-dimensional 
polyhedron $P=\{x:Hx\leq b\}$  given  
 via the matrix $H$ and column $b$ as follows:
\[  H=\left[ \begin{array}{ccc}
-1 & 0    & 0              \\
 0 &  -1    & 0              \\
 1  &  0   &  0            \\
 0 &  1    &  0             \\
 1  & 0    & -1    \\ 
 0 &  1    & -1    \\      
-1 &  0   &  -1    \\   
 0 &  -1  & -1        
              \end{array}  \right] , \y
     b=\left[ \begin{array}{c}
         1  \\ 1  \\1  \\1  \\1  \\1  \\1  \\1  
\end{array}  \right] . \]

Assuming no prior information on the set $P$, the natural way to proceed  
is as follows.   
It has first to be ascertained whether the quest for a VR-representation 
makes sense and is non-trivial. To see whether this is the case 
first call {\tt initv=initvert(H,b)}.  
This routine does  one of the
following things: it emits a message {\tt polyhedron Hx<=b has no vertex},
or a message {\tt polyhedron Hx<=b is empty}, in these two cases  
returning with an empty matrix {\tt initv}; or it displays  a message 
{\tt  vertex found } and a 2-column matrix {\tt  initv }. The first column 
of  {\tt initv}
gives a vertex of the polyhedron, the second column the indices at which 
the vertex satisfies the inequality system with equality. 
Only in this third  case it makes sense to determine the VR-representation.     
Since, in theory, in very badly behaving 
cases {\tt initvert} might take a long time to come up with an answer - if 
$H$ is an $m\times d$ matrix there may be ${m\choose d}$ submatrices to check -
you may possibly also see a message 
{\tt Systematic vertex search begins. This may take time}, displayed 
whenever the routine
switches from a probabilistic to a deterministic search procedure. 

However, believing that users of this implementation 
wish to examine not too large polyhedra  only  occasionally, we hope this will 
be no serious     
obstacle. At any rate, users owning the \Matlab Optimization toolbox 
can optionally resort  
to a judicious use \Matlab'\nny s linear programming routine {\tt lp.m}  to find 
an initial vertex; or they can adapt our simplex routine {\tt simplex.m}, 
using the ideas  described in [Sch, p131c5].

\parbox{4cm}{ 
 \begin{tabular}{cccc} 
      {\tt initv}&=& {\tt -1} &{\tt 1 }  \\ 
                 & & {\tt -1} &{\tt 2 }  \\
                 & & {\tt  0} &{\tt 7 } 
              \end{tabular} }\y
\parbox{12.5cm}{Assume now you have found via the call to {\tt initvert.m}, or otherwise, a vertex 
and herewith a nonempty  2-column matrix {\tt initv}.  In the example under consideration,
it is the matrix on the left. }

The first column indicates a vertex, and the second column the row indices defining 
a regular $3\times 3$ submatrix of $H;$ and at the same time indices of inequalities 
satisfied at the vertex with equality.  Now call   
{\tt [V,R]=htovr(H,b,initv)} . 
Type {\tt V} and {\tt R}.  You get 
\begin{verbatim}
V=  0     1     1    -1    -1      R=  1     0     1     0    -1     0    -1     0
    0     1    -1     1    -1          1     0    -1     0     1     0    -1     0
   -1     0     0     0     0;         0     1     0     1     0     1     0     1;
\end{verbatim}
Matrix {\tt V} saves the vertices of the polyhedron as columns; {\tt R} the 
rays in the form origin (of ray)/ray, origin/ray, etc. In the present case, 
the last four vertices  originate the ray 
{\tt [0 0 1]'}.   \halmos

The file {\tt htovr.m} finds the VR-representation via
two lines of code: first it  calls \\
{\tt [Dictionary, Basis, Nbasis]=tratodic(H,b,initv).}
This translates the triple {\tt (H, b, initv)} to a  
 $(m+1)\,\times\, (m+d+2)$-matrix, 
  {\tt Dictionary}, which is the initial dictionary and also furnishes initial ordered sets  
{\tt Basis} and {\tt Nbasis},  about which more later.                   
Then it calls 
{\tt [V,R]=lrs(Dictionary, Basis, Nbasis)}.  This yields the desired vertices
and extreme rays. 

As emerges,      {\it lrs} is at heart a  
H-to-VR-transform; however,  
a standard  lifting technique, see the explanations for {\tt vtohr.m} in 
section 3, allows to solve with {\it lrs} also  the converse problem.

{\bf 1.3 Example } From a VR-representation to a H-representation.  \y
We suppose known a VR-representation of a $d$-dimensional pointed polyhedron.
The rays should be given without information concerning their origins.      
Assume, say, 
we begin with the set $V$ and adapted $R$ we have found in example 1. So  
\begin{verbatim}
V=[ 0     1     1    -1    -1         R=[ 0
    0     1    -1     1    -1             0 
   -1     0     0     0     0];           1]; 
\end{verbatim}
With this input, a simple call to {\tt [H,b]=vrtoh(V,R)} yields  up to multiplication   
with 2 the following matrix $H$ and column $b.$

\begin{verbatim}
H=-1     0     0             b=   1
           0    -1     0                  1
           0     1     0                  1
           1     0     0                  1
          -1     0    -1                  1
           0    -1    -1                  1
           0     1    -1                  1
           1     0    -1                  1 ,
\end{verbatim}
or an equivalent system. This is equivalent to what we started with in example 1.2.   \halmos

We have not yet have made extensive use of our implementation, but  
close this section with preliminarily reporting that a VR-to-H translation  of 
the 13 dimensional polytope of 24 vertices, occurring
naturally in the study of the so called Oliveira-Marcus conjecture for 
$4\times 4$-matrices, see 
[D, p254c-1] and [K] has according to our {\it lrs} implementation 
82 facets: 16 facets with 18 vertices,  48 facets with  13  vertices,  
                                    18 facets with 20 vertices. 
Furthermore every vertex is contained in exactly 53 of the facets. 
A more detailed exposition on this polytope will be available  soon.  \\

{\bf 2. How does the \Matlab implementation of {\it lrs} work ?}

The ideas behind {\it lrs} are well explained in Avis' paper [A]. It 
is assumed the reader is familiar with that paper and has it at hand. This section is 
rather to mediate between Avis' paper and the \Matlab implementation adopted. 
Users unfamiliar with \Matlab should first read the short section 3. \\

\parbox{16cm}{
\begin{center}
\vspace*{-0.5cm} \epsfxsize = 8.75cm  \epsfbox{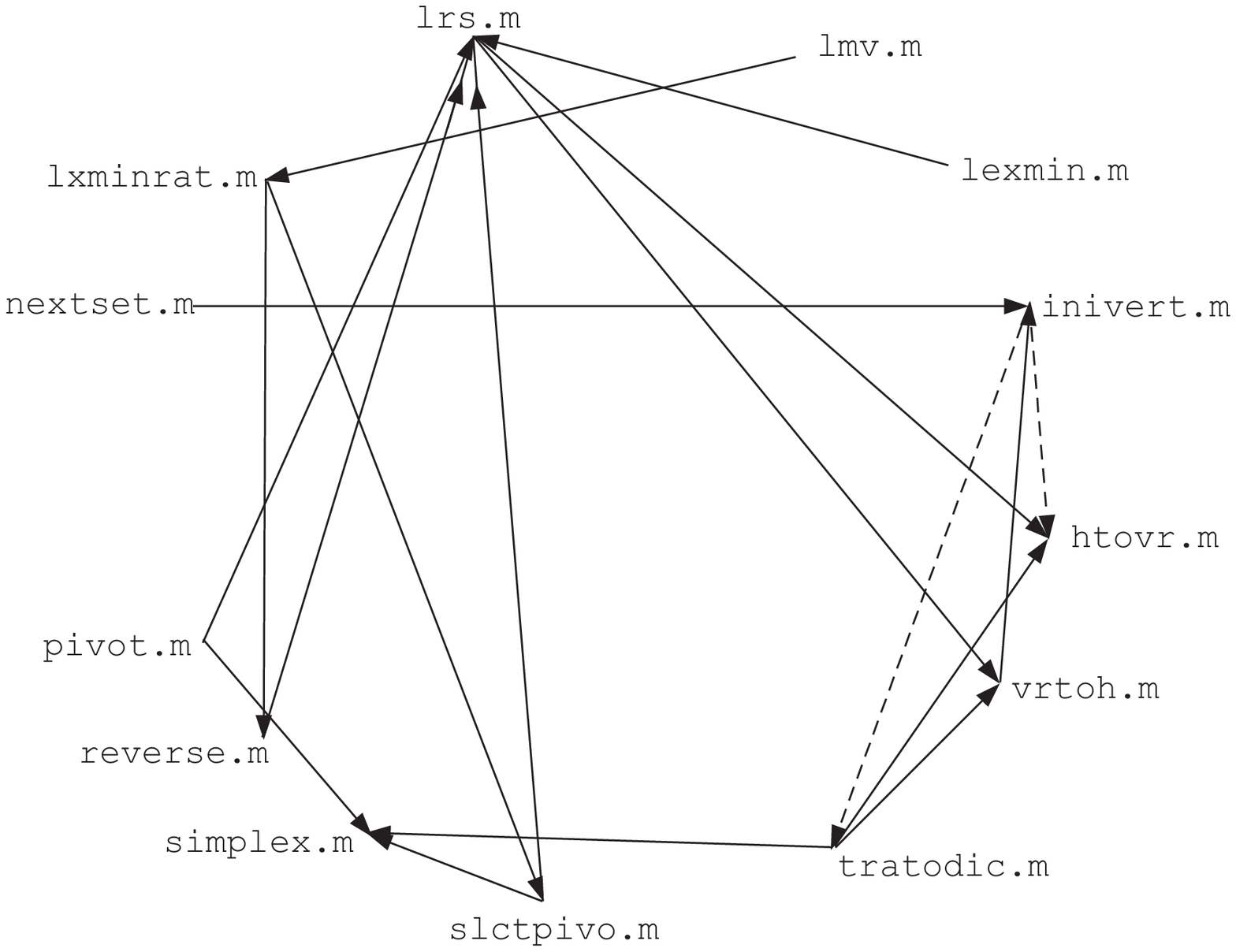}
\end{center}
%\vspace{12cm}
}

\vspace*{0.25cm}
{\bf 2.1 Figure} 
The {\it lrs}-implementation uses 12  M-files; one further file, 
{\tt simplex.m}, implements the simplex algorithm and was written  because 
we could take advantage of the fact that
{\it lrs} required to write almost all  routines that also occur as 
ingredients in the simplex algorithm of linear programming.  The files 
{\tt lexmin.m}, {\tt lxminrat.m}, {\tt pivot.m}, 
{\tt reverse.m}, {\tt slctpivo.m}, implement  the functions 
named in [A] almost identically, but recall the DOS operating system allows 
only eight-letter names.  
The principal file is {\tt lrs.m}. It implements the orchestration given  
by the pseudo-code [A, p192]. The files {\tt lmv.m}, {\tt nextset.m} are auxiliary, while
the remaining files where already mentioned in section 1.  
These files are  arranged alphabetically 
in the above diagramm in counterclockwise order beginning with 
{\tt htovr.m}. An arrow, e.g.  {\tt nextset.m} $\ra${\tt initvert.m},
means that the function {\tt initvert.m} calls upon {\tt nextset.m}. 
If no arrow points towards  a  function,  it 
depends solely on standard \Matlab\nny.  \halmos

As mentioned in the previous section, the {\it lrs}-algorithm 
does not start at the H-description of 
a pointed  polyhedron itself. The H-description must first be translated into an 
initial dictionary. This is done via {\tt tratodic.m} which uses the 
result of {\tt initvert.m} which calls {\tt nextset.m}. 
So, referring to the codes of these routines, it is natural to give now the  
correctness proofs of these functions in this order. 
 All code is given 
in section 4; the heads of those files, also shown when calling  
{\tt help file-name },
give hypotheses and conclusions of what
is to be proved. 
Concerning notation, below $1_n$ is, depending on context, the row 
or column with $n$ ones; $0_n$ is similar, $x$ can mean a row or column.
As dogmatic coherence seems impossible and undesirable to achieve, we change between 
typewriter- and \TeX fonts in hopefully reasonably a manner.
We also use a notation we mimick from \Matlab\nny. If 
$x=(x_1,x_2,\ldots  )$ and $i,j\in \ZZ_{>0}^n$, we write 
$x_{i:j}=(x_i,\ldots, x_j).$    Notations like $x_{d+1:d+m}$ are to be understood
in accordance with \Matlab's precedence rules as $x_{(d+1):(d+m)},$ etc.

{\tt nextset.m} correctness proof: \\ 
It is easy to see from the definition of lexicographic ordering
of the family of $k$-subsets of an $n$-set, that two 
$\{0,1\}$-$n$-tuples  encoding $k$-subsets of $\{1,\ldots,n\}$ are immediate
successors of each other if they  have 
the respective forms  $[*,\underline{1,0}, 0_{m_0}, 1_{m_1}],$ 
and $[*,\underline{0,1}, 1_{m_1}, 0_{m_0}].$  
Here $*$ represents the same initial segments and the zeros or ones can be partially 
absent, as happens e.g. for $n=4$   
if $k=0$ or $k=4,$ or, if $k=2,$ with the first and last 4-tuples 
[1 1 0 0] and [0 0 1 1].
Our program works as follows. In line 2, {\tt last0} gives the index of the rightmost
$0$ in {\tt x}; {\tt xinit} is the initial segment of {\tt x} strictly to the left  of 
this $0.$ Strictly to the right of the 0 we find a block of final 1s that extends 
till the end of {\tt x}. This block is saved in {\tt xfin1s}. In $j$ defined 
in line 3, we find the index of 
the rightmost 1 in {\tt xinit}. If such a 1 does not exist, then {\tt x} is of 
the form $[0,\mbox{(zeros)},0,1,\mbox{(ones)},1],$   which means it is the 
lexicographically last element (in the class considered). In this case, as claimed,
{\tt x} itself is returned and execution terminates via the {\tt return} in line 4.  
Otherwise, we begin constructing {\tt nextx} in line 5,
by putting {\tt nextx=x} and modifying the present {\tt [1 0]} 
in entries $j,j+1$  (underlined above)
of {\tt x} to {\tt [0 1]}. Directly after this, we place                
the block {\tt xfin1s} and fill in line 6 the remaining final entries with 0es. 
One also can easily verify that the code does what it should if applied to 
special cases that yield  {\tt xinit} or {\tt xfin1s} as empty vectors.
\halmos

{\tt initvert.m} correctness proof: \\ 
We have the following fact; see [Sch, p99c-6\& p104c-7]:
a set $P=\{x: Hx\leq b\}$ is either empty or a nonempty polyhedron whose 
minimal faces have dimension $n-$rank$(H).$ \y
Now, a vertex is a minimal face  of dimension 0. 
It follows that, {\tt  if rank(H)<n } happens in line 2, the correct message is 
transmitted, and the function terminates via {\tt  return}. 
In the opposite 
case we have   $m\geq n=\mbox{rank}(H)\geq 1 $ and  the problem is to find 
a regular $n\times n$ submatrix  $T$ of $H$ such that the solution $x_0=${\tt x0} to 
the subsystem of equations obtained from $Hx\leq b$ by satisfying the 
corresponding inequalities with equality,  also satisfies $H x_0 \leq b.$ 
To find such a (vertex) $x_0$ and corresponding indices satisfied with equality, 
we first try briefly a  probabilistic, then, if unlucky, a deterministic method.

After entering the lines 4 to 18 {\tt while} -loop, we enter 
line 5 at most {\tt cl} times defining there a random $\{0,1\}$-$m$-row {\tt z} 
with $n$ 1s and $m-n$ 0s, increasing  variable {\tt cnt}, originally set 
{\tt 1}, each time we execute it. While {\tt cnt}$\leq${\tt cl}, we jump to line 9 and 
define there a submatrix {\tt  T}  of {\tt H}. 
If the choice of {\tt z} was lucky {\tt  T} has rank {\tt n} and defines 
a vertex. The code of lines 10 to 13 shows that in this case the vertex 
and the set {\tt I} indexing rows for corresponding equations is returned; 
we will exit via the {\tt return} of line 12. During this
phase line 14   guarantees after each  
unlucky trial a jump back to line 5.

%-p6beg

After the {\tt  cl}-th unlucky search we enter line 6 and begin a systematic 
vertex search to reach a definitive answer: 
we define  the $m-$tuple $z=[1_n, 0_{m-n}]$ 
which is  (the characteristic function of) the lexicographically 
first $n$-set in $\{1,2,\ldots, m\}.$ 
The code sets   {\tt cnt=cl+2}, a value  never changed later. 
Thus in all subsequent processing, commands in lines 5 to 8 remain unexecuted,          
while those of lines 14 to 17 gain importance; in particular
the value of {\tt z} is updated 
via {\tt z=nextset(z)} in  line 16 with each execution of the loop.  
Recall that  {\tt z=nextset(z)} gives the
lexicographically next $\{0,1\}$-$m$-tuple with exactly $n$ entries 1, or, 
if we have arrived at the end, then $z=[0_{m-n},1_n]$ is reproduced.  
This latter {\tt z} is the  unique one  among the considered $m$-tuples, whose last 
$n$ entries are all equal to 1; that is for which  the  statements in the 
{\tt if } of line 15 are exectuted. This shows that we leave the
{\tt while } loop  via one of the following events: \y
Case: {\tt return}  is encountered in line 12 within the body of the {\tt if} of
lines 10 to 13. In this case the code shows us that we 
have encountered an $n\times n$-submatrix $T$ of $H$ so that {\tt rank(T)=n}  
(is true) and the unique
$x$ solving $T x=${\tt b(z)} is up to a tolerance of 10 machine epsilons 
such that $b-Hx\geq 0,$  that is $Hx\leq b.$ 
In other words, 
$x$ is a vertex. In that case message and output is  correct. 
(On the used machine,    {\tt eps}$\approx 2.3 \cdot 10^{-16}.$)
\y
Case: the  {\tt return} of line 12 is not encountered: since   
rank$(H)=n,$ 
there must be an $n\times n$-submatrix of rank $n.$ The preceding discussion
shows  that the ranks of all $n\times n$
submatrices of $H$ are examined. Hence 
the result in [Sch] cited allows    to conclude that $P$ is empty. The    
function terminates via the {\tt return} of line 15. In this case, the
message  is in agreement with truth again, and, by their initialisation,
$x_0,$ and $I$ will remain empty, as  claimed. \halmos

Note: The probabilistic phase turned out useful in particular in 
{\tt initvert.m}-calls from {\tt vrtoh.m}. The reasons are explained in the 
latter's correctness proof.

{\tt tratodic.m} correctness proof: \\ 
The first step in {\tt lrs} as well as in the simplex algorithm is a 
\underline{tra}nslation 
of the system of inequalities describing the polyhedron and the linear objective
function in\underline{to} a {\it\underline{dic}tionary}. Certain expositions,  
e.g. [Sch], use for this or a closely related concept the designation `tableau'.

Suppose we are given a pointed nonempty polyhedron $P$ 
described by a linear system of inequalities:  $P=\{x: Hx\leq b\}.$  We assume
$H$ is $m\times d,$ hence $x$ is a real $d$-column, and $b$ a real $m$-column: 
$H\in \RR^{m\times d}, x\in \RR^{d\times 1}, b\in \RR^{m\times 1}.$
Given $Hx_{1:d} \leq b_{1:m},$ we can find $m$ nonnegative numbers collected in an
$m$-column $x_{d+1:d+m}$ of {\it slack variables} such that $$Hx_{1:d}+x_{d+1:d+m} =b. \eqno(1)$$  
So $P=\{x_{1:d}: Hx_{1:d}+x_{d+1:d+m}=b, x_{d+1:d+m}\geq 0 \}.$
We know already  that $x_{1:d}$ represents a vertex of $P$ iff 
we find $d$ of the inequalities $Hx_{1:d}\leq b_{1:m}$ 
corresponding to linearly independent rows of $H$  satisfied with equality. 
In other words $d$ of the variables   $x_{d+1:m+d}$ corresponding to 
linearly independent rows are $0.$  We fix an initial vertex and 
follow Avis' development [A, p182c6] in assuming after possibly
reordering the rows of $H$ that this vertex is determined by putting the final 
$d$ slack variables to 0:  $x_{m+1:m+d}=0.$   We write 
$H_{\rm upp},$ $H_{\rm low}$  for the upper $m-d$ and lower $d$ rows of (reordered) 
$H$ respectively. 
Then $H_{\rm low}$ is invertible. Using also   $b_{\rm upp}=b_{1:m-d},$ 
and $b_{\rm low}=b_{m-d+1:m},$ 
equation (1) splits into two subsystems 
\begin{tabbing}
xxxxxxxxxxxxxxxxxxxxxxxxxxxxxx\=xxxxxxxxxxxxxxxxxxxxxxxxxxxxxxxxxxxxxxxxxxxxxxx\=xxxxxxxxxxx \kill
 \>\y $ H_{\rm upp} x_{1:d} +x_{d+1:m} = b_{\rm upp}, $    \> $(2_{\rm upp})$   \\
 \>$\; H_{\rm low} x_{1:d} +x_{m+1:m+d} = b_{\rm low}.$  \> $(2_{\rm low})$
\end{tabbing} 
%$$\begin{array}{rclr}
% H_{\rm upp} x_{1:d} +x_{d+1:m}&=&b_{\rm upp}.   &\y (2_{\rm upp})   \\
% H_{\rm low} x_{1:d} +x_{m+1:m+d}&=&b_{\rm low}  &\y (2_{\rm low})
% \end{array} $$
From the lower system, we obtain
$x_{1:d}+H_{\rm low}^{-1} x_{m+1:m+d}=H_{\rm low}^{-1} b_{\rm low}.$
Plugging the resulting $x_{1:d}$  into the upper system yields 
$x_{d+1:m} + H_{\rm upp} (-H_{\rm low}^{-1} x_{m+1:m+d}  
         +H_{\rm low}^{-1} b_{\rm low})  =b_{\rm upp},$  or
$x_{d+1:m} - H_{\rm upp} H_{\rm low}^{-1} x_{m+1:m+d}=b_{\rm upp}-H_{\rm upp} H_{\rm low}^{-1} b_{\rm low}$ 
These facts show that the original system is equivalent to 
$$ \left[ \begin{array}{c|c}
 \y I_d \y 0_{d\times (m-d)}         & H_{\rm low}^{-1}          \\ 
 0_{(m-d)\times d} \hy I_{m-d}  &-H_{\rm upp}H_{\rm low}^{-1}
              \end{array}  \right]x_{1:m+d} =   
\left[ \begin{array}{c}
     H_{\rm low}^{-1} b_{\rm low}    \\ 
     b_{\rm upp}-H_{\rm upp}H_{\rm low}^{-1}b_{\rm low}
              \end{array}  \right].  \eqno(3)   $$
Note that the matrix on the left of the vertical line is $I_m.$
By definition, the initial dictionary  associated to the system $Hx\leq b$ 
is now the $(m+1)\times (m+d+2)$ matrix obtained as follows:  
first write the augmented matrix of the 
above system, that is, include the right hand side column; second modify
(border) this latter matrix so as to accomodate the objective function [A, p182c10],
%We opted to restrict the file tratodic.m to Avis' objective value $x_0$  
given  by $x_0+1_d x_{m+1:m+d}=0.$ This means 
joining to the left of the augmented matrix referred a zero-column  
and finally putting on top of the  resulting matrix the row
$[1,\, 0_m, \, 1_d,\, 0].$  We can now analyse the code.

After determining the size of $H$ in  line 1 as $m\times d,$ 
line 2 defines $I$ as an indicator $\{0,1\}$-$m$-column having $d$ 1's exactly at 
the positions that the entries of {\tt initv(:,2)} indicate. The new matrix  
$H$ and column $b$ produced correspond to a rearrangement of the system 
according to the suggestion that the final $d$ slack variables put to 0
should be the inital vertex.  Line 3 produces the upper and lower  
parts of matrix $H$ and column $b;$ in line 4 we define 
{\tt invHlow}$=H_{\rm low}^{-1}$.  The {\tt Dictionary} of line 5 encodes 
precisely the augmented matrix of the linear system (2). 
In line 6 the $m\times (m+d+1)$   matrix of line 5 is modified to an 
$(m+1)\times (m+d+2)$-matrix in order to accomodate Avis' objective row. 
This matrix is actually precisely the matrix  $[A|b]$ corresponding
to  [A, p182c-4], would we have presented the polytope $P$ using his settings 
[p180c-0].  So Avis' initial dictionary and the {\tt Dictionary} 
produced by {\tt tratodic.m} differ only in that we have included the right 
hand side column. 
Line 7 implements a heuristic also implemented in {\tt pivot.m}.
It is supposed that all entries of the dictionary that are smaller than 
100 machine epsilons 
are actually meant to be 0. For an - albeit superficial - justification of this, 
see the notes after the proof to {\tt pivot.m}.
With line 8 {\tt tratodic.m} exits with the initial 
{\tt Dictionary}, {\tt Basis}, and {\tt Nbasis}.  \halmos

{\bf 2.2 Example} If we call {\tt initv=initvert(H,b)} with $H,b$ as below (representing  
a heavily perturbed, origin-centered, axes-parallel unit cube with side lengths 2),
we get the shown matrix {\tt initv}. 
\begin{verbatim}
H=[1.0  0.1   -0.3             b=[0.9      initv=1.018    1
  -1.0  0.4   -0.2                1.1            1.22     3
   0    1.0    0.1                1.3            0.8      5
   0   -1.0    0.7                0.8                       
   0    0      1.0                0.8                       
   0    0.1   -1.0];              1.2];                     
\end{verbatim}
Now, calling 
{\tt [Dictionary, Basis,Nbasis]=tratodic(H,b,initv)}, yields 
\begin{verbatim}
Dictionary=
    1    0    0    0    0    0    0    1       1       1      0
    0    1    0    0    0    0    0    1.0    -0.1   0.31   1.018 
    0    0    1    0    0    0    0    0.0     1.0  -0.1    1.22  
    0    0    0    1    0    0    0    0.0     0.0   1.0    0.8
    0    0    0    0    1    0    0    1.0    -0.5   0.55   1.79  
    0    0    0    0    0    1    0    0.0     1.0  -0.8    1.46  
    0    0    0    0    0    0    1    0.0    -0.1   1.01   1.878 

Basis= 1  2  3  4  5  6  7 ,  Nbasis = 8  9  10.
\end{verbatim}
This is but one example of a family of dictionaries, bases  
and cobases or non-bases, namely they are the corresponding {\it initial } objects. 
Invoking {\tt lrs.m} or {\tt simplex.m} does a number of pivot operations 
via {\tt pivot.m}  that constantly 
manipulate {\tt Dictionary}, {\tt Basis}, and {\tt Nbasis}. 

{\bf 2.3 Observation} At any stage, these three objects satisfy the following:\\
i. The {\tt Basis} is an $(m+1)$-row such that {\tt Basis(1:d+1)}$=[1,2,\ldots,d+1].$ \\
ii.    {\tt Nbasis} is a  $d-$row satisfying          
       {\tt Nbasis}$\sbs\{d+2,\ldots,m+d+1\}.$\\

%-p8beg

iii.    As sets, {\tt Basis} and {\tt Nbasis} partition the set $\{1,\ldots, m+d+1\};$  
   as ordered tuples they are usually {\it not} ascendingly ordered  
   although   they are so at start.  \\
iv.     {\tt Dictionary} is an $(m+1)\times (m+d+2)$-matrix such that
      {\tt Dictionary(:,Basis)}$=I_{m+1},$ the $(m+1)\times (m+1)$ identity matrix.
    All dictionaries are  row equivalent to each other and in particular 
            to the initial dictionary using only multiplication of a row with  
            a nonzero scalar and addition of a scalar multiple of one row to another as 
            elementary row operations;  i.e. no interchanges of rows are used. \\
v. A triple {\tt (Dictionary, Basis, Nbasis)}   
has the property that each of its entries determines the other: if 
$(D,B,N)$ and $(D',B',N')$ are triples of that type with $B=B',$ then necessarily 
$D=D'$ and $N=N',$ etc. \\
vi.  If $[D'|b']$ is any matrix that is 
            row equivalent to a dictionary, then the  polyhedron $P$ can be described 
            as  the set of points $x_{2:d+1} \in \RR^d$
such that $x=(x_1,\ldots, x_{m+d+1})$ satisfies $D'x=b'$  and  $x_{d+2:m+d+1}\geq 0,$ i.o.w.,
      $$ P=\{x_{2:d+1}: D'x=b' \mbox{ and } x_{d+2:m+d+1}\geq 0 \}.$$
vii.     Among all such matrices $[D'|b']$ the dictionaries  
 allow to extract information on extremals(=vertices and extreme rays) 
in a particular convenient way. Namely, if {\tt Dictionary}$=[D|b],$ 
{\tt Basis}, and {\tt Nbasis} is a triple as in v, then a solution $x$ to the system  
$D{\tt x}={\tt b}, {\tt x}\geq 0, {\tt x(Nbasis)}=0,$  restricts to a 
vertex $x_{2:d+1}=${\tt x(2:d+1)}. In fact, since $Dx=b$ is equivalent to 
{\tt Dictionary(:,Basis)*x(Basis)+} \\{\tt Dictionary(:,Nbasis)*x(Nbasis)=b}, we see by iv
that {\tt x(Basis)=b} and, by i, the vertex is given by 
{\tt x(2:d+1)=b(2:d+1)=Dictionary(2:d+1,sD2)}; see below for {\tt sD1, sD2}.     
To see the relation of this to equation (3) above keep in mind that 
most indices here used would be by 1 smaller,  would the  
first row of {\tt Dictionary} not express the objective function and    
that \Matlab 
does not allow adresses to matrix entries to be $0$ or negative,  forcing  
us to deviate slightly from Avis' indexations.  \\
viii. The column $A_B^{-1}A_s$ occuring frequently in [A] is just {\tt Dictionary(:,s)},
as follows from  p184c-6. \\
The proofs for i to vii  follow from  the proofs for 
{\tt tratodic.m} and {\tt pivot.m} (below). \halmos

Note: As follows from these explanations, all the dictionaries, bases, and 
cobases have the size of the respective initial objects. While for simplicity 
of code and(?)  speed, the variables {\tt Dictionary}, 
{\tt Basis}, {\tt Nbasis}, are declared global by the functions where they
are first used, namely {\tt lrs.m} and {\tt simplex.m},  
and the functions these call, the size of {\tt Dictionary} and the 
cardinality (length) of {\tt Nbasis} are redetermined 
in each function necessary, via the commando {\tt [sD1,sD2]=size(Dictionary)}. 
Hence we always have ${\tt sD1}=m+1=|{\tt Basis}| $; ${\tt sD2}=m+d+2$; and conversely, 
$m={\tt sD1}-1,$ 
$d={\tt sD2}-{\tt sD1}-1=|{\tt Nbasis}|={\tt lNb}$ (=\underline{{\tt l}}ength of 
{\tt \underline{N}basis}).

{\bf 2.4  Example} (=2.1 continued). Upon calling {\tt pivot(r,s)} for an {\tt r$\in$Basis} and an 
{\tt s$\in$Nbasis}, 
we produce a linear system row equivalent to the previous having at column {\tt s} 
the {\tt r}-th standard vector. To illustrate,
if we do on the previous example  {\tt pivot(5,8)}, followed {\tt pivot(7,10)},  
we get 
\begin{verbatim}
Dictionary =
  1  0  0  0 -1  0   -0.4455  0    1.5446  0   -2.6267
  0  1  0  0 -1  0    0.2376  0    0.3762  0   -0.3257
  0  0  1  0  0  0    0.0990  0    0.9901  0    1.4059
  0  0  0  1  0  0   -0.9901  0    0.0990  0   -1.0594 
  0  0  0  0  1  0   -0.5446  1   -0.4455  0    0.7673  
  0  0  0  0  0  1    0.7921  0    0.9208  0    2.9475
  0  0  0  0  0  0    0.9901  0   -0.0990  1    1.8594 
Basis= 1  2  3  4  8  6  10 ,  Nbasis = 5  9  7.
\end{verbatim}
\halmos 

To ascertain these claims this is a good place to present the 

{\tt pivot.m} correctness proof: \\
%In the note to the  correctness proof of tratodic it 
%was claimed that in any stage {\tt Dictionary(:, Basis)+ is $I_{m+1}.$  
%We prove this here.
%Basis is an {\it ordered} set, indeed it is always a {\tt sD1}-tuple of the form
% {\tt [1:d+1 ...]}.
An $r\in {\tt Basis}$ occupies a position, say $t,$ found in line 4,  
as is there found the position {\tt tn}  of  $s\in {\tt Nbasis}.$ In other words,
{\tt r=Basis(t)}, {\tt s=NBasis(tn)}. So                  
{\tt Dictionary(:,r)} is a column that is the $t$-th standard vector 
(i.e. has a single 1 in $t$-th entry, 0s elsewhere). By line 3, line 5            
divides the t-th row by the value that stands in the Dictionary's $s-$th column. 
The command  {\tt Dictionary(t,s)=1} there is to guarantee a clean 1. 
The loop lines 6-8  cleans all other entries of column $s$ to 0,
and yields a new dictionary equivalent to the entering one. In 
particular {\tt Dictionary(:,s)} is now the t-th standard vector. 
In line 9 the appropriate updatings are done: by line 4, element $s$ in the 
cobasis and $r$ in the basis are interchanged leaving all other positions  
unchanged; so the `conservation law' 
${\tt Basis}\uplus{\tt NBasis}=\{1,2,\ldots, m+d+1\}$ is obeyed.     
A heuristic is used to account for 
numerical errors: we assumed  that 
values in modulus $\leq 10^{-14}$ are meant to be zero and did say so explicitly 
in the line  10. To go sure, line 11 also produces 
a clean standard vector for column $s$ of the dictionary.  \halmos

Notes: 
As mentioned,  numerical errors and their propagation are the 
Achilles-heel of the implementation.
The effect of lines 10, 11 can be `fabulous' (?).
In the example of the polytope referred at the end of section 1, 
in which {\tt pivot.m}, as explained here            
led to the system of 82 inequalities, the supression of these lines led,
apparently due to blow up of numerical noise, to some 780 inequalities.

{\bf 2.5 Observation} For understanding the mechanics of {\it lrs} 
as well as of the simplex algorithm, it is important to understand 
a polyhedron and motions in it via dictionaries. 
If we base ourselves on the description given in vi above, i.e.
$P=\{x_{2:d+1}: D'x=b' \mbox{ and } x_{d+2:m+d+1}\geq 0 \},$
then  we home in at a vertex of $P$ if we    put $d$ of 
those nonnegative variables 
in $x_{d+2:m+d+1}$  equal to 0, that could appear in an {\tt Nbasis}.
To leave that vertex along an edge or ray, means to increase exactly one 
of these variables, say $x_s,$ $s\in{\tt Nbasis},$ while counterbalancing this 
increase
by means of the variables in {\tt Basis} so as to satisfy the system of 
equations. If we assume $[D'|b']$ to be an accompanying {\tt Dictionary}, we 
see that this means that $x_{2:d+1}$ `moves' exactly along the 
opposite direction of the subtuple with components {\tt 2:d+1} given in column $s$
of the dictionary. 
So {\tt -Dictionary(2:d+1,s)} is the direction of an 
edge or extremal ray of the polyhedron incident at  the vertex 
{\tt Dictionary(2:d+1,sD2)} under consideration. Can we distinguish 
between an edge and a ray? We are considering a ray by definition iff we 
could increase $x_s$ indefinitely and yet always select  $x_{2:d+1}$ 
within $P.$ We see that this is true if and only if 
{\tt Dictionary(2:d+1,s)}$\leq 0_{d-1}.$ In the above example 2.4 this is not the
case for any $s\in {\tt Nbasis}$ since the perturbed cube does not allow 
extremal rays; but we also see that to the possible values of $s$ correspond 
three directions that are roughly those incident at the vertex $(-1, 1, -1)$
of the origin-centered axes-parallel unit cube with sides 2. Call a vertex 
of a $d-$ dimensional polytope {\it simple } if it is contained in 
exactly $d$ facets; or equivalently if it is incident with exactly $d$ 1-dimensional
faces (=edges or extreme rays). A given dictionary can always exhibit only 
$d$ such faces; and  a non-simple vertex allows more than one associated 
dictionary. {\it Simple } polytopes are those that (like the cube) have 
only simple vertices. We hope
these observations neatly illustrate [A, p183c-8] and other explanations 
in section 1 of [A]. \halmos

Our next major aim is to explain the implementation of the simplex algorithm 
for the reason that it contains many ingredients also occuring in the more   
complicated {\it lrs}-algorithm. 
In accordance with figure 2.1, the necessary files  will   be explained 
in the logical order {\tt lmv.m}, {\tt lxminrat.m},  {\tt slctpivo.m}, 
{\tt simplex.m}. 

{\tt lmv.m} correctness proof: \\
If {\tt M} is an empty matrix or a matrix with only one row, then
the lines 2-4-{\tt while}-loop is never entered and the output is either ${\tt I}=\emptyset$ or 
${\tt I}=\{1\}$ respectively, and hence correct. Now assume {\tt M} to 
be $m\times n$ with $m\geq 2.$ Then the while-loop is entered. Consider initiating its
body with any set  $I\sbs I_0=\{1,\ldots, m\},$ $I\neq  \emptyset.$ 
Then in line 3,  {\tt m} holds the minimum 
value of column {\tt j} of matrix {\tt M(I,:)}; 
and {\tt sI} holds the position of indices  of entries equal to {\tt m} 
{\it within} the
ordered set {\tt I}. So we find  $\emptyset \neq {\tt I(sI)}\sbs {\tt I}$ 
is such that 
{\tt M(I(sI),j)} is a subcolumn of  {\tt M(I,j)}, all whose entries are equal to $m,$
while all other entries (if any) of {\tt M(I,j)} are strictly larger than 
{\tt m}. 
We also note that if $|{\tt I}|=1,$ then {\tt I(sI)=I}. \y Let $I_j$ be
the value of {\tt I} after $j\geq 1$ executions of the while's body. 
Clearly the while loop is executed a total of $e\leq n$ times. 
The arguments just given and a look at the termination condition for the  
while loop permit us to say this: we have \y
a. $I_0\sps I_1\sps I_2 \sps \ldots \sps I_e\neq \emptyset;$ \hy
b. $e\geq 1$ and $|I_e|=1$ or $e=n$; 
c. For any $j=1,\ldots, e$: ${\tt M}(I_j,{\tt j})$ is a subcolumn of 
${\tt M}(I_{j-1},{\tt j})$ all whose 
entries are equal a certain $m_j,$ while  
all the entries in ${\tt M}(I_{j-1}\setminus I_j,{\tt j})$  
(if any) are$>m_j.$  \y
d. The program with the while's entering condition simplified to just 
    ${\tt j}\leq n$ would yield the same output (but take possibly longer). 
\y We can and will now assume the simplified  enter condition mentioned in d, and hence have $e=n.$
Since by a, $I_n\sbs I_j,$ we find that ${\tt M}(I_n,{\tt j})=m_j$, for 
${\tt j}=1,\ldots, n.$
So ${\tt M}(I_n,:)$ consists of $|I_n|$ rows, each equal to 
$\hat{m}=[m_1,\ldots, m_n].$ Let
$\hat{r}=[r_1,\ldots, r_n]$ be any of the rows of ${\tt M}.$  By 
c, $m_1 \leq r_1.$ If $m_1<r_1,$ then $\hat{m}<_{\rm lex}\hat{r}.$ If 
$m_1=r_1,$ then $\hat{r}$ is by c part of ${\tt M}(I_1,{\tt :}).$ By c, $m_2\leq$any entry in 
${\tt M}(I_1,{\tt 2}).$ Hence $m_2\leq r_2.$ If $m_2<r_2,$ then 
$\hat{m}<_{\rm lex} \hat{r}.$ 
If $m_2=r_2,$ then $\hat{r}$ is  part of ${\tt M}(I_2,{\tt :}).$ Repeating this argument,
we see that the outputted ${\tt I},$ being equal to $I_n,$ is as claimed. \halmos

{\tt lxminrat.m}  correctness proof: \\ 
As usual, line 1 guarantees the function has access to the current 
dictionary, basis, and cobasis, while line 2 determines  the sizes of 
these objects. So because of [A, p184c-4 to p185c2], 
line 6  defines the matrix $D$ there referred; and line 3
yields Avis' $a=A_B^{-1} A_s$ (see observation 2.3viii). Next, in line 4,
the set of candidate indices $i$ of [A, p185c8] are determined. 
The added term {\tt lNb+1} is necessary since otherwise we would  get the 
indices relative to the short column {\tt a(lNb+2:sD1)}.
If {\tt I} is empty, 
then [A, p185c-8] tells us to return with {\tt r=0}. This is done in line 5. 
The discussion also shows that {\tt I} is empty iff {\tt s} represents a ray.
So control flow encounters line 7 if and only if  ${\tt I}\neq \emptyset.$  
In this case matrix {\tt Dtilde} collects the rows referred in 
[A, p185c8].
Since {\tt D}  has full row rank, the set defined by {\tt i} defined in 
line 8 consists of  only one element. Note that {\tt i} is relative to the number of rows os 
{\tt Dtilde}. But {\tt t=I(i)} 
will give index $t$ such that $D^t/a_t$ is lexicographically minimum 
in Avis' p185c7 sense. Finally, line 9   tells us the basis element 
at the position $t;$  i.o.w. {\tt Dictionary(:,r)} is the $t$-th standard vector. 
Note also that {\tt Basis} does not contain 0. We can conclude that the output  
{\tt r} is $0$ iff $s$ represents 
a ray.  \halmos

{\tt slctpivo.m} correctness proof:   \\
Line 1 opens access to {\tt Dictionary}, {\tt Basis}, and {\tt Nbasis}. 
Assuming for a moment {\tt Dictionary} nonoptimal, its first row has negative 
elements; see [A, p186c8] and the explanations to {\tt simplex.m} below.
In this case line 2 will define a natural {\tt s}$\leq${\tt sD2-1}.  Since by 
observation 2.3 iv, {\tt Dictionary(1,Basis)}$=[1, 0_m],$ we find
 ${\tt s}\in {\tt Nbasis}.$  Hence {\tt j}
in line 3 takes an integer value so that {\tt Nbasis(j)=s}, and hence 
{\tt Dictionary(1, Nbasis(j))}
is among the negative reals occuring in the dictionary's first row,  
the leftmost one. The line 5 assures return of the value  of {\tt r} 
desired by the specifications [A, p191c6\& 185c10]. 
If the dictionary is optimal, then  {\tt s},{\tt j} are   
{\tt []},{\tt []}. So the {\tt return} in line 4 will 
assure we exit with {\tt r=0, j=[]}.  \halmos

As mentioned, {\it lrs} uses ideas of the simplex method. 
Indeed, as a byproduct of the files we had to write for {\it lrs}, all essential 
elements for implementing the simplex-algorithm with the lexicographic pivot-rule
are available. For this reason we decided to add that algorithm itself as a 
benefit. The reader unfamiliar with the theory behind the simplex method should 
consult  first, e.g.,
Schrijver [Sch] pages 129-131c5, followed by p132-133c5. He should read those  
explanations without assuming in his equation (11) that $x\geq 0.$ Equation  
(12) is not of real importance either. This means to 
assume his matrix 
$  \left[ \begin{array}{c}
   -I     \\
   A     
     \end{array}  \right]  $
as an object used for referencing in (13). Also,  the matrices $C_k$ there used,
are always submatrices of $A$.      With this the conclusions concerning 
$u_k$ in [Sch, p133c3] remain valid. The explanations on p133 concerning the 
simplex tableau remain valid if one begins  in the origin 
as Schrijver does in his phase II where he supposes $b\geq 0.$ In our  
more general case however, the origin will usually not be a vertex of $P$ 
and hence we have to assume the usually  nonzero value {\tt u(sD2-sD1)} 
in the right upper entry of
the initial  {\tt Dictionary}; see explanations that follow. \\

{\tt simplex.m}  correctness proof: \\
Line 1 declares {\tt Dictionary}, {\tt Basis}, and {\tt Nbasis}
to be global in order the functions {\tt pivot} and {\tt slctpivo} have access
to these `variables' after they have been produced by a call to {\tt tratodic.m} in line 2.
Recall that Avis' $x_0$ is our $x_1$.
Assume we have to maximize $x_1=c_1x_2+\ldots +c_dx_{d+1}=cx_{2:d+1}$. Recall that
the initial dictionary as delivered by {\tt tratodic.m}  has the form 
{\tt [$I_{m+1}$ |Dictionary(:,sD1+1:sD2-1) |b]}. We need to modify its row 1,
the objective row, in accordance with the {\tt c} given.  
% maximization desired in {\tt simplex.m} it has the wrong objective row.  
{\tt Dictionary} encodes in rows 2 to $d+1$ the   
equations $x_{2:d+1}+${\tt Dictionary(2:d+1,sD1+1:sD2-1)}$x_{m+1:m+d}=b_{2:d+1}.$
Multiplying the corresponding matrix equation from the left with the row {\tt c} 
and keeping the note after observation 2.3 in mind,
we obtain             
$x_1+${\tt  c*Dictionary(2:lNb+1,sD1+1:sD2-1)}$x_{m+1:m+d}=cb_{2:d+1}.$
Now, at the initial vertex $x_{m+1:m+d}=0$. So, having defined {\tt u} as 
in line 4, 
$x_1=cx_{2:d+1}=cb_{2:d+1}=$\\{\tt c*Dictionary(2:d+1,sD2)=u(sD2-sD1)}, 
and we see that the correct objective row is now given as defined in line 5.
\hy Next we select in line 6 a pair {\tt [r,j]}
according to the lexicographic pivot rule. The specifications of   
{\tt slctpivo.m} tell us that {\tt j=[]}  iff there is no negative value 
in the first row. In this case simplex-theory says that the right  
upper corner entry of {\tt Dictionary } is the optimum.
Consider now we  obtain a nonempty {\tt j}. Then {\tt s=Nbasis(j)} makes  
sense. If the column $a=${\tt Dictionary(:,s)} has from the $d+1$-st entry onwards 
only nonpositive values, we  
get $r=0$ and know by the discussion in Avis p182c-7 and p185c13, and observation 2.5,  
that 
we have an extreme ray. Moving along that away from its origin, by the pivot rule, 
we obtain arbitrary large objective values.  Otherwise, $r\neq 0,$
and we can do a pivot to obtain a  value of the objective function not 
smaller than the previous one. The reason is that in [A, p185c-4]
we have $\bar{D}^0=D^0-a_0 \bar{D}^t,$  $a_0<0$  and due to the lex-positivity
of the bases in lex-positive pivoting, we add a nonnegative quantity to the 
uppermost entry of the last column of the dictionary.  We do this via 
the {\tt while}-loop of lines 7 to 11 various times. The theory of the 
simplex algorithm  
guarantees that we will either exit {\tt simplex.m} via line 8 or obtain in the 
upper right corner of {\tt Dictionary} the maximum value {\tt maxim} of the 
objective function on the polytope. Lines 13 and 12 will give us that value
and a vertex at which it is obtained.   \halmos

{\bf Observation 2.6} We review rapidly the philosophy of {\it lrs}. 
Let  $x_0$ be a fixed vertex of $P$  and let $B^*=\{1,2,\ldots,m+1\}.$ Then there
exists a dictionary exhibiting $x_0$ as initial vertex, $B^*$ as basis, and 
encoding in its first row an objective function that assumes in $x_0$ its maximum:
indeed, provided {\tt initv} is encodes $x_0$, {\tt tratodic(H,b,initv)} produces 
this dictionary.  Consider the family of all dictionaries equivalent to this    
one and associated with lex-positive basis. 
As Avis theorem p187c8 explains, these dictionaries cover all vertices 
of $P$ and if we apply the simplex algorithm to any one of these,  that is if 
we would run the code piece lines 6 to 13 of {\tt simplex.m}, we generate 
finite paths of lex-positive dictionaries. 
The union of these paths evidently defines a tree whose nodes are 
lex-positive dictionaries equivalent to its root, 
the dictionary initially chosen for $x_0$ as common end point. 

Upon demanding {\tt x(Nbasis)=0}, we can see any sequence of triples {\tt (Dictionary, Basis, Nbasis)} 
as a sequence of visits to vertices and certain of the edges or extreme rays
incident at it. Geometrically explained, {\it lrs }  visits vertices by running 
in a depth-first manner  through that tree,    
beginning at the vertex determined by the initial dictionary  as its root.
To do so, half of the steps in {\it lrs } are done in the directions 
that are exactly the reverse ones the simplex algorithm would do. 
We  prove the correctness of the file {\tt reverse.m} choosing these 
steps. Note that
the corresponding proposition 6.1 [A, p192c-3] is marred by a number of mistakes: the 
corrected pieces should read as follows: ... for any $t=0,1,...,m$ let ... 
and only if (i) $w_v^0>0,$ ...  by the pivot formula $\bar{w}_u=-w_v^0/a_i$...
... ratio test for each positive coefficient ...  \halmos

{\tt reverse.m} correctness proof: \\
The correctness proof is based on the definition of 
{\tt reverse} as given on [A, p191c-6]  and its equivalent characterization  
established in (corrected) proposition 6.1. What 6.1 precisely says is that  
we should get {\tt brev=1} iff there hold  6.1i, 6.1ii, i.e.  
{\tt u=lexminrat(v)~= 0} (i.e. {\tt v} represents an edge, not a ray), and, 
for the then existing {\tt i} with {\tt u=Basis(i)}, also 6.1iii.  
Further, in case {\tt brev=1} this {\tt u} should be returned.    We establish 
that precisely this is the outcome of the function. 

Indeed if 6.1ii does not hold then the lines 2,3 
will guarantee that {\tt brev=0} is returned and no more lines
are executed.  In the opposite case 6.1ii holds.  Under this assumption we 
have to show that {\tt brev=1} or {\tt brev=0}  is delivered according 
to whether or not 6.1i$\&$iii is satisfied.
Indeed if     6.1ii  holds, line 2               
delivers an {\tt u} in {\tt Basis}. For such an {\tt u}, {\tt u}$\neq 0.$ 
So line 3 is skipped, and line 4 will be successfully executed.
Consider what happens in  line 7 and the two lines preceding it. 
Evidently {\tt wbar} is precisely the $\bar{w}$ of 6.1iii.
Set  {\tt J} of line 8 holds all elements $j$ of {\tt Nbasis} with    $j<u.$  
By 6.1iii we have to check
whether all entries  of {\tt wbar} indexed by elements $j\in J$ are $\geq 0.$ 
If $J$ is empty this is trivially true and we have nothing to check: the 
boolean variable {\tt wbarg0} in line 9 is put equal to 1. 
If $J\neq \emptyset,$ then {\tt  wbar(J)>=-10*eps} is a meaningful  
expression involving a relational operator {\tt >=}.
It is  a $\{0,1\}$-row  with 1s in all entries where {\tt wbar(J)}  
has entries in practice$\geq 0$ and 0s where the values are $<0.$ 
Hence {\tt wbarg0} has value 1/0 according to whether/not the condition 6.1iii holds.  
We see from lines 9,10 that {\tt brev=1} and the correct {\tt u} will be 
returned iff condition 6.1i$\&$iii is satisfied.    \halmos

The final ingredients before proving the correctness of  {\it lrs }
proper is a proof of the correctness 
of {\tt lexmin.m} determining the cases in which a vertex or 
a ray should be outputted: recall p189c-0, according to which there exists 
a 1-1 correspondence in general not  between the bases and vertices and rays, but
between the lex-min bases and these objects.

{\tt lexmin.m}  correctness proof:  \\
Upon leaving   line 2, we have in {\tt b} the last column of {\tt Dictionary}, 
and {\tt boolean=1}. Upon calling {\tt lexmin(0)},  one wishes to decide whether 
the current {\tt Basis} is lex-minimal for the current basic feasible solution 
(associated to the vertex, see [Sch, p134c-3]).
{\tt boolean=1/0} is to be outputted according to if/not {\tt Basis} is lex-minimal. 
Evidently lines 3 to 7 are executed. According to 
[A, proposition 5.1, p187], we 
have to put {\tt boolean=0}  iff we can find {\tt r$\in$Basis}, 
{\tt s$\in$Nbasis}, with $r>s$
such that the current basic feasible solution exhibits $x_r=0,$ and (with 
subindex corrected to $t$), 
$(A_B^{-1}A_s)_t\neq 0.$    (The idea of that proof is that if such objects 
exist, a {\tt pivot(r,s)} would yield the same vertex, but a smaller basis.)
Now a pair $(r,s)=(${\tt Basis(t)}$,s)$, as measured after line 5, satisfies 
evidently $r>s.$ Recall that `to be at the vertex' means putting {\tt x(Nbasis)}$=0.$  
In view of  {\tt Dictionary(:,Basis)}$=I_{m+1},$ saying $x_r=0,$ is seen to be
equivalent to saying that {\tt b(t)}$=0.$  From observation 2.3viii it is now 
clear that  line 6 puts {\tt boolean=0} iff it should do so.  \y 
Upon calling {\tt lexmin(v)} from {\tt lrs.m}, we call it  with {\tt v$\in$Nbasis}
representing a ray, and want to   decide whether the basis is lex-minimal 
for this ray. This time we should put {\tt boolean=0}  iff conditions i to iv
of proposition 5.4 on p189 are satisfied. Evidently, we execute this time lines 8 to 12.
The similarity of this proposition with 5.1 and 5.4
is reflected in that of the code. It needs no more explanation except saying that
the additional condition in line 11, namely {\tt Dictionary(t,v)=0}, implements 
condition 5.4iv.  \halmos

Note: the comparisons with exact 0 (in 
{\tt Dictionary(t,v)==0}, etc.) should not cause trouble because of the numerical 
noise eliminations in {\tt tratodic.m} and  {\tt pivot.m}. Other functions 
do not alter  dictionary entries. 

{\tt lrs.m} correctness proof: \\
Line 1 clears global variables that might have survived
from previous uses of lrs. Line 2 declares {\tt Dictionary}, 
{\tt Basis}, {\tt Nbasis} 
global, in order to make them  accessible to functions invoked by {\tt lrs.m}, 
without formal transference. Line 3 determines the format of {\tt Dictionary}, 
puts {\tt j=1},  and uses the note after  observation 2.3, 
according to which $|{\tt Nbasis}|$={\tt lNb}. 

The remainder of the code is encapsulated in the 
line 4 to 22 {\tt  while 1 ... end} loop. 
This loop is begun by the lines 5 to 17 {\tt  while j<=lNb ... end}
loop, which     is certainly entered.
Since upon entering,  1$\leq${\tt j}$\leq${\tt lNb}, line 6 will always 
give a well-defined result for {\tt v$\in$Nbasis}. In turn we get in line 7 a  
well defined tuple {\tt [brev,u]$\in \{0,1\}\times (\{0\}\cup$Basis}).

A look at lines 8,12 shows that  either the code piece lines 9, 10   or the 
code piece lines 13, 14, 15 will be tried to for execution. 

{\it Claim:} The lines 5 to 17 {\tt while} loop will terminate. $\lhalm$
Since we augment  $j$ whenever ${\tt brev}=0$, nontermination implies that we have infinitely 
often ${\tt brev}=1$, and hence do infinitely many pivots at line 13. Since there are only 
finitely many bases, this means that we cycle. But by line 7   the pivots are 
done always such that they could be reversed via the lexicographic pivot rule: 
applying to a basis $B$ the sequence of commands written in lines 7,13, yields 
a new basis to which applying lines 18,20 would yield $B$ again. We see that
cycling would allow the same for pivoting via the lexicographic pivot rule.  
But this rule is known to render cycling impossible [A, p186c-7]. 
Hence the claim.  $\rhalm$

Consider `measuring' the sequence of  pairs {\tt (brev,j)} between lines 16 and 17. 
This sequence will have the following form 
$(0,2),(0,3),\ldots, (0,j_0)$, $(1,1),(0,2),\ldots,(0,j_1),$ 
$(1,1),(0,2),....,(0,j_2), ..., (1,1),(0,2),$ 
$...,(0,j_k)=${\tt (0,lNb+1)}, whereupon we leave the 5-17-loop. 
Some of the sequences $(0,2),\ldots,(0,j_s)$ may be missing. Associated with 
each $(1,1)$ is a pivot yielding a dictionary that was not encountered before 
within the loop. Every such pivot corresponds to a step deeper into the tree,  
approaching a leaf. Note  that we have done $k$ pivots and so $k$ base changes.

Let $B,B'=B-u+v$ be the $(k-1)$-st and $k$-th bases. We see 
{\tt B'=pivot(u,Nbasis(}$j_{k-1}$)).
As we exit with a certain dictionary and {\tt (brev,j)}$=(0,j_k)$ at line 17,  
line 18 gives us a pair {\tt [r,j]} such that: either {\tt j=[]}, which 
means that we are inspecting the optimal dictionary and  line 19 
will  guarantee that we return to the function calling {\it lrs}; or: we 
do in line 20 a pivot returning to the $(k-1)$-st of the bases produced above. 
In that case line 18 yielded {\tt [r,j]==[r,$j_{k-1}$]} as follows from 
proposition 4.1b, p185. Line 21 guarantees that upon entering line 5, we will 
examine the {\tt Dictionary} associated to $B$  
from a column  not yet examined onwards.   

Although vertices, as indicated by various dictionaries, are usually revisited,   
it will be clear by now that a pair \hy {\tt Dictionary}/column {\tt Nbasis(j)}, \hy 
once examined, is never revisited.  
Furthermore, as the
lexicographic rule leads from lex-positive to lex-positive dictionaries, only 
lex-positive dictionaries are visited. 

{\it Claim:} All lex-positive dictionaries are visited. 
$\lhalm$  If we are given any lex-positive dictionary $D$, the simplex algorithm 
with lexicographic pivot selection and adequate objective row
`climbs' from $D$ to the, by proposition [A, p187c1] unique,  lex-positive dictionary 
with basis $B^*$ in $x_0.$ This means that there is also a way back to $D$
which  along the run of {\it lrs } is  indicated by  {\tt reverse.m}. $\rhalm$

The  claim just proved, and the facts that lex-minimal dictionaries are lex-positive, 
[A, p188c1\&c-7],  
guarantee that we visit all lex-minimal dictionaries. Since all vertices and rays 
have lex-min representations, the specifications of {\tt reverse.m} and lines 9 and 14
guarantee that all vertices and all origin/ray pairs are outputted exactly once.     \halmos

The last pieces missing are the correctness proofs for the input output 
routines {\tt htovr.m } and {\tt vrtoh.m}. 

{\tt htovr.m } correctness proof: \\
The correctness of this two lines code 
follows directly from the specifications of {\tt tratodic.m}, {\tt lrs.m}, 
and {\tt initvert.m}. \halmos

The explanation for file {\tt vrtoh.m}  needs some preparatory remarks. 

Recall that {\it lrs } is at heart a method for solving H-to-VR-problems: given 
an H-representation of a pointed polyhedron, it finds a VR-representation. Just as often 
however we need to go in the inverse direction. 
So assume a polyhedron $P$ is given by a nonempty set $V$ of vertices and rays $R$, that is,  
$P$ is given in VR-representation, and we want to find its H-representation. 
This VR-to-H-problem can be solved by  solving a related
H-to-VR-problem,  and therefore falls within the realm of {\it lrs}. The ideas are 
as follows.            

Let $P$ be   $d$-dimensional.  We lift it first to a cone: each vertex $(a_1,\ldots, a_d)$ 
is lifted to a point  $(1, a_1,\ldots, a_d)$ in $(d+1)$-space, identifiable with a ray, 
as shown by the figure borrowed from [Z, p31], also showing the cone $C(P)$ 
generated by these. 

%\includegraphics[keepaspectrario,scale =0.36]{graphic2}

%\hspace{8cm} \parbox{4cm}{}
\begin{center}
\epsfxsize= 7cm \epsfbox{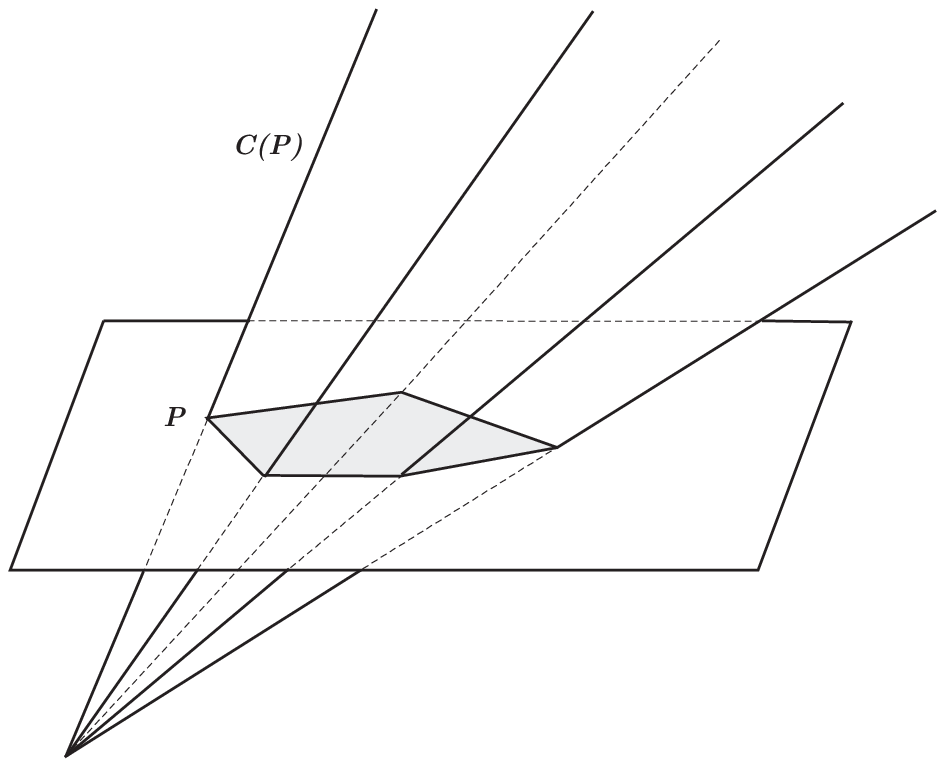}
\end{center}

{If $P$ is  unbounded then, in
addition to vertices, we  have  rays in $P$ itself. A ray in $P$ with 
direction $(a_1,\ldots, a_d)$ is naturally 
identified with the ray $(0,a_1,\ldots, a_d)$ in $C(P).$
\y The cone $C(P)$ and its polar $C(P)^\Delta=\{c: cx\leq 0 \;\mbox{for all $x\in C(P)$}\}$
determine each other. Each ray $(1, a_1,\ldots, a_d)$ of $C(P)$ defines  
its perpendicular plane through the origin; the halfspace it delimits is given by 
the inequality   
$-x_1-a_1x_2-\ldots-a_{d}x_{d+1} \leq 0$ which determines a facet of $C(P)^\Delta.$ 
Similarly to ray directions $(0,a_1,\ldots, a_d)$ 
of $P$ there correspond the inequalities $-a_1x_2-\ldots-a_{d}x_{d+1} \leq 0.$
In this way we have obtained from the VR-description of $P$ a H-description
of $C(P)^\Delta.$ }

Applying {\it lrs} to this H-description of $C(P)^\Delta,$ we get a 
VR-description of $C(P)^\Delta;$ of course with the only vertex 0. The rays of 
$C(P)^\Delta$ correspond to facets in $C(P)=(C(P)^\Delta )^\Delta; $
see [Z, p64c7\& p62c4]. 

These remarks and the proof below will  suffice to illuminate the correspondences 
mentioned in in [A, p194c-2 to 195c7] and used in the file {\tt vrtoh.m}.  Note that 
Avis' $\bar{P}=C(P)^\Delta$ above.

{\tt vrtoh.m} correctness proof:\\
 {\tt V} and {\tt R} are given as columns. Line 1       
generates a matrix {\tt E} with 
columns consisting of the vectors from vertex in column 1 to other vertices.
To avoid column 1 of  {\tt E } becoming the zero column it is skipped via 
{\tt E(:,1)=[]} and the rays are added.
In other words, \verb+[E,R]+ 
represents the polytope+char.cone description mentioned in Schrijver p106c-2.
The rank of {\tt E} therefore is equal 
to the affine dimension of the  convex polyhedron generated by {\tt V} and 
{\tt R}. An appropriate error message is issued in lines 2 to 4, if 
the polyhedron is not  full dimensional and further processing is aborted.  

Otherwise the polyhedron is full dimensional and
lines  5,  6, 7  do the translations described
in [A, p195c1to3] to a system of linear inequalities $A x\leq 0$ in one higher 
dimension: a H-description of  pointed cone $\bar{P}.$
Hence it makes sense to search for 
a quadratic submatrix  $A_0$ of $A$ of rank equal to the number, say $n$ of 
columns of $A.$ 
This is done via the 
{\tt  initv=initvert(A,b)} in line 8. The fact that we deal with a  cone 
having  the origin $0_n$  as (the only) vertex implies that the vertex search is 
actually equivalent to finding $A_0$, for  $A0_n\leq 0$ is trivially satisfied.
It is here that the probabilistic phase in  
{\tt initvert} finds an appropriate $A_0$  
sometimes much faster than the deterministic does. 
With this the input is prepared for invoking {\tt tratodic.m} in the next line 9,
whose output is in turn transferred to be processed with  {\it lrs}. 
The output of this is a 
vertex-ray representation of $\bar{P}.$ $\bar{P}$ being a pointed cone, again
{\tt V} consists of the origin only.   The translation of the 
rays has to be done. Recall that the {\tt R} in line 10 obtained as output 
of  running {\it lrs} gives us geometric information on 
the starting point of the rays in the form 
origin/ray,  origin/ray, etc.                    
The fact that only the even column indices of {\tt R} are in our case of 
interest  is recognized via the collection of rays given in line 11.
In the last line the column collection of 
rays is transformed in a row collection and into a system of inequalities 
$Hx\leq b$
of the original dimension, as prescribed by [A, p195c4]: to a ray 
$(z_1,\ldots, z_{d+1})$ of $\bar{P}$ there corresponds the inequality
$-z_2x_1-\ldots -z_{d+1} x_d \leq z_1$.  \halmos \\

{\bf 3. Some \Matlab basics }

Here are some specifics of the \Matlab  programming language, in order to  
facilitate understanding the code for those having no experience with \Matlab\nny.
Since things are tried to be kept simple and brief,  
the explanations are necessarily somewhat superficial.
Our \Matlab routines are functions,  saved as usual in so-called 
M-files with the extension {\tt .m},  called  precisely via the name of 
the function: 

\parbox{5cm}{ 
\begin{tabular}{l} 
{\tt function m=addone(n)} \\
{\tt m=n+1;} 
 \end{tabular} } 
\parbox{12cm}{A file, say {\tt addone.m}, with the content on the left
can be called from within another function writing a line like 
{\tt k=addone(7)}, yielding $k=8.$ } 

The fundamental object in \Matlab  are matrices with complex or real entries. 
Rows or  columns,  as vectors, are to be considered $1\times n$ or $n\times 1$ matrices
respectively; 
scalars as $1\times 1$ matrices. Matrices {\tt A,B} of appropriate sizes  can be added, subtracted,  
multiplied, or entrywise multiplied by commands {\tt A+B}, {\tt A-B}, {\tt A*B}, {\tt A.*B}, 
respectively.
If {\tt r} is a scalar, then {\tt r+B} adds {\tt r} to each entry of {\tt B}, 
while {\tt r*B} multiplies each entry of 
{\tt B} by {\tt r}. Transposition of {\tt A} is given by {\tt A'}.

\parbox{5cm}{
\begin{tabular}{l}  
{\tt A=[ 1  -7  5   13 } \\ 
{\tt \y \hhy  -2 \hhy 4  3    7 } \\ 
{\tt \hy \hhy -19  11 8   2]; } \\                                     
{\tt I=[2 2 1]; J=[3 4];} \\
\end{tabular}   

yields

\begin{tabular}{l}
{\tt  A(I,J)=3     7 } \\  
{\tt  \y\y\y  3     7 }  \\
{\tt  \y\y\y  5    13. } \\
\end{tabular} }
\hy\parbox{11.5cm}{
Matlab variables have no type declarations.  Matrices can 
be entered as on the left, or in fashions like {\tt A=[2 1; 4 -5]; }. The 
 `{\tt ;}' within the brackets separates the rows, so we created a $2\times 2$ 
 matrix $A;$  the second `{\tt ;}' supresses printing of {\tt A}. 
Let {\tt A} be an $m\times n$ matrix. Assume {\tt I, J} are rows or columns 
consisting entirely of integers selected from $\{1,2,\ldots, m\}$ and
$\{1,2,\ldots, n\}.$  Then {\tt A(I,J)} is an $|I|\times |J|$ matrix containing 
the rows and columns of {\tt A} in the order and number they appear in {\tt I} and {\tt J}. 
See the example at the left. } 

Writing  {\tt 3:7} isolatedly is the same as {\tt [3 4 5 6 7]}, while 
{\tt 7:-2:1} is {\tt [7 5 3 1]}, but be careful with precedences:  
{\tt j+k:l+m} is the same as {\tt (j+k):(l+m)}, so {\tt 1+2:4} is 
{\tt [3 4]}, while {\tt 1+[2:4]} or {\tt 1+(2:4)} is 
{\tt [3 4 5]}.
Genuine submatrices can also be indicated via characteristic vectors.
Thus {\tt A(1:2, 2:4)} is the same as {\tt A([1 1 0], [0 1 1 1])}. 
Commands like 
{\tt A(:,J)}  or  {\tt A(I,:)} indicate that all rows or columns respectively 
have to be included; so if {\tt A} is $m\times n,$ they mean the same as 
{\tt A(1:m,J), A(I,1:n)}. 

Relational operators {\tt <,<=,>,>=,==,~=}, perform element-by-element
comparison between two matrices of the same size or with a   scalar,
yielding 0 or 1 according 
to whether the relation is false or true. 
The result of {\tt A==-19} using the matrix above is a $3\times 4$ matrix 
with a single 1 in position $(3,1),$ and 0es elsewhere. 
Logical operators \verb# &, |, ~# represent and, or, negation, respectively. 
They work element-wise on matrices of same 
size, or if one of these is a  scalar. A {\tt 0} 
at input of a logical expression represents false, anything nonzero 
represents true.
The output of a logical expression is a $\{0,1\}$-matrix of appropriate size. 

Structures such as {\tt if } {\sf condition} ... {\tt end}, 
{\tt while} {\sf condition} ... {\tt end}, {\tt for}  
{\sf variable=expression}  ... {\tt end} 
work largely as elsewhere;  \Matlab allows conveniences. For example
{\tt for j=[1:3 5:-1:3] ... end}  means we wish to take {\tt j} the 
values 1,2,3, 5,4,3,
in this order.   In a syntactically valid construct
{\tt if} {\sf condition} {\sf statements} {\tt end},
the  {\sf condition} evaluates to a matrix. If the real parts of that 
matrix are all nonzero the statements are executed, otherwise not. 
For example the command within {\tt if [1 0 1]  j=7; end}  is not executed, 
but in {\tt if [.1 1 -5] j=7; end} it is executed. 
Similar behaviour holds for the {\tt while}.

We explain briefly the standard \Matlab functions we used. In the 
following paragraph {\tt A,B} represent matrices, {\tt x} a vector.

The command
{\tt [m,n]=size(A)}  yields the size of {\tt A} as $m\times n$, 
{\tt rank(A)}  its rank;  {\tt inv(A)} is the 
inverse of {\tt A}, and  {\tt abs(A)} the 
matrix obtained by taking the absolute values of matrix {\tt A}. 
To solve an equation $Ax=b$ for $x,$ given appropriately sized $A,b,$ type 
\verb#x=A\b#. Commands
{\tt ones(m,n)} and {\tt zeros(m,n)} yield the 
$m\times n$ matrices consisting entirely of ones and zeros respectively,  
while {\tt eye(m)} is the $m\times m$ identity matrix. 
Command {\tt isempty(A)} yields 
1 or 0 according to if or not {\tt A} is an empty matrix.  
{\tt 1./A} is a matrix of the 
same size having  at address (i,j) the entry {\tt 1/A(i,j)}, showing the 
infinity symbol {\tt Inf} wherever $A(i,j)=0.$
The command {\tt diag(x)}
creates a square matrix having {\tt x} in the diagonal; 
the command 
{\tt length}(x)
yields the number of entries of  {\tt x}; {\tt max(x)}  the maximum element
of {\tt x},  {\tt find(x)} yields the indices of the nonzero entries of 
{\tt x}, {\tt all(x)} returns 1 if all of the elements of a vector {\tt x} 
are nonzero, {\tt sum(x)} is the sum of the elements of  {\tt x},  
{\tt randperm(m)} generates a random permutation on ${\tt 1,\ldots, {\tt m}}.$

\parbox{5cm}{
\begin{tabular}{l}
{\tt function foo }\\
{\tt j=0; disp('Here I am');}  \\
{\tt disp('Now I am here');} \\
{\tt while j<100000 j=j+1; end} 
\end{tabular} } \y
\parbox{11.5cm}{ The command {\tt disp('text')} can be used to display  a string {\tt text}. 
While developing the program, we used it
at points to know which function is executed at a given moment. We had to discover, 
it can deceive:  in a file like at the left
the second string {\tt Now I am here} would not be 
displayed until after the {\tt while} is processed.} 

This phenomenon can give 
at times the wrong impression of a suspended execution.   A {\tt pause(1)} before  
the {\tt disp} seems to resolve such problems. 

A {\tt return} causes a normal return to the invoking function or keyboard,
a {\tt clear } or a {\tt clear global} clears all or only all global 
variables: such a command in  
a function  avoids that  a global variable {\tt D}  created in previous of its 
uses is  unintentionally used; e.g.  extended in 
constructions like {\tt D=[D,d]}, etc.  Variables not declared {\tt global} 
are as usual only locally accessible to the function where they are created. 
To be accessible to a function called by another one, a variable 
has to be declared {\tt global} in both of the functions.  \\

{\bf References}

\parbox{1cm}{[A] \\}\hy \parbox{15.5cm}{D. Avis, A Revised Implementation of the Reverse Search Vertex Enumeration Algorithm, in [KZ, 177-197].} 

\parbox{1cm}{[D]\\}\hy \parbox{15.5cm}{S. W. Drury, A Counterexample to a Question of Merkikoski and Virtanen on the 
 Compounds of Unitary Matrices, {\it Linear Algebra Appl.} 168:251-257(1992). }

\parbox{1cm}{[K]\\}\hy \parbox{15.5cm}{A. Kovacec, The Marcus-de Oliveira conjecture, bilinear forms, and cones, 
{\it Linear Algebra Appl.} 289:243-259(1999).  }          

\parbox{1cm}{[KZ]\\}\hy \parbox{15.5cm}{G. Kalai, G. Ziegler (ed.): Polytopes - Combinatorics and Computation, 
 DMV Seminar 29, Birkh\"auser, 2000. ISBN 3-7643-6351-7. }

\parbox{1cm}{[M]\\ \\}\hy \parbox{15.5cm}{N. Myhrvold (Vice-president of Microsoft):  `Software is a gas; 
it expands to fill its container. After all, if we hadn't 
brought your processor to its knees, why else would you get a new one?', 
{\it Scientific American}, July 1997,  page 69, bottom. } 

\parbox{1cm}{[Mat]}\hy \parbox{15.5cm}{\Matlab 3.1,  The MathWorks, Inc. 1992. }

\parbox{1cm}{[Z]}\hy \parbox{15.5cm}{G. M. Ziegler: Lectures on Polytopes, GTM 152, Springer 1998.}

\vspace{1cm}

NOTE: Matlab-code next pages. 

\pagebreak

{\bf 4. \Matlab-Code}

This section collects the code of the 13 M-files, alphabetically ordered; for 
use delete the line numbers {\tt 1., 2.}, etc. and put each function 
in a separate M-file preferrentially in the same directory of the \Matlab path.

\begin{verbatim}
function [V,R]=htovr(H,b,initv)
% [V,R]=htovr(H,b,initv)
% in: matrices H,b, and  vertex-information as produced by initv=initvert(H,b). 
% out: a collection of vertices and  rays with information about their origin. 
1. [Dictionary, Basis, Nbasis]=tratodic(H,b,initv);
2. [V,R]=lrs(Dictionary, Basis, Nbasis);
----------
function initv=initvert(H,b)
% initv=initvert(H,b) in: an mxn-matrix H and an m-column  b. 
% out: if exists, a pair initv=[x0,I] of vertex x0 of the polyhedron Hx<=b;
%    and a n-tuple I of positive integers  so that   H(I,:) has rank n 
%    and H(I,:)*x0=b; otherwise x0=[], I=[], and appropriate messages.
   
 1.  [m,n]=size(H); x0=[]; I=[];
 2.  if rank(H)<n  disp('polyhedron Hx<=b has no vertex'); return; end
 3.  cnt=1; cl=10;
 
 4. while 1
 5.  if cnt<=cl   p=randperm(m); p=p(1:n); z=zeros(1,m); z(p)=ones(1,n); cnt=cnt+1; end
 6.  if cnt==cl+1   z=[ones(1,n), zeros(1,m-n)];  cnt=cl+2; 
 7.   disp('Systematic vertex search begins. This may take time.'); 
 8.  end
 9.   T=H(z,:);             
10.  if rank(T)==n 
11.   x=T\(b(z));  u=b-H*x; f=(u>=-10*eps); 
12.   if all(f) x0=x; I=z; disp('vertex found');  initv=[x0, (find(I))'];  return; end 
13.  end
14.   if cnt==cl+2
15.    if z(m-n+1:m)==1  disp('polyhedron Hx<=b is empty'); return; end    
16.    z=nextset(z); 
17.   end
18. end
----------
function  boolean=lexmin(v)
% boolean=lexmin: in: v=0 or v in Nbasis. out: boolean=0 or boolean=1.                    
% if v=0: boolean=1 iff Basis is lexmin for a basic feasible solution. 
% if v in Nbasis: boolean=1 iff Basis is lexmin for a geometric ray represented by v.
 
 1. global Dictionary Basis Nbasis 
 2. [sD1,sD2]=size(Dictionary);  b=Dictionary(:,sD2); boolean=1;     
 3. if v==0
 4. for s=Nbasis
 5.  for  t=find(Basis>s)     
 6.    if  b(t)==0 & Dictionary(t,s)~=0   boolean=0; return;  
 7.    end, end, end, end
 
 8. if ~(v==0)
 9. for s=Nbasis
10.  for  t=find(Basis>s)
11.     if  b(t)==0 & Dictionary(t,s)~=0 & Dictionary(t,v)==0 boolean=0; return;  
12.     end, end, end, end
----------
\end{verbatim}

\begin{verbatim}
function I=lmv(M)
% I=lmv(M) finds, given a real matrix M indices of rows that are 
% lexicographically smallest. 
 
 1. j=1; I=1:size(M,1); n=size(M,2);
 2. while (length(I)>=2)&(j<=n)
 3.   col=M(I,j);   m=min(col);  sI=find(col==m);  I=I(sI); j=j+1;
 4. end
-----------
function  [vertices, rays]=lrs(Dictionary,Basis,Nbasis) 
% [vertices, rays]=lrs(Dictionary,Basis,Nbasis): given a valid triple (D,B,N) as
% input, as obtained from tratodic.m, returns all vertices and extreme rays 
% of associated polytope in  `vertices' and `rays'  in form `origin ray origin ray ... '. 
 
 1. clear global  
 2. global Dictionary Basis Nbasis 
 3. [sD1,sD2]=size(Dictionary); j=1;  lNb=sD2-sD1-1;     

 4. while 1 
 5.   while j<=lNb
 6.     v=Nbasis(j);    
 7.     [brev, u]=reverse(v);
        
 8.     if ~brev         
 9.      if u==0&lexmin(v) rays=[rays, Dictionary(2:lNb+1,sD2) -Dictionary(2:lNb+1, v)]; end 
10.      j=j+1;
11.     end
        
12.     if brev
13.       pivot(u,v);
14.       if lexmin(0)  vertices=[vertices, Dictionary(2:lNb+1,sD2)];  end
15.       j=1; 
16.     end
17.   end  
       
18.  [r,j]=slctpivo; 

19.   if isempty(j) vertices=[vertices,Dictionary(2:lNb+1,sD2)]; return; end   
20.   pivot(r,Nbasis(j)); 
21.   j=j+1;
22. end
---------
\end{verbatim}

\begin{verbatim}
function r=lxminrat(s)
% r=lxminrat(s): given s in Nbasis and lexpositive Basis calculates integer
% r=lexminratio(Basis,s) in sense of [A, p185c6...11].  
% in particular r==0 iff s represents a ray, an r in Basis otherwise.
 
 1. global Dictionary Basis Nbasis
 2. [sD1, sD2]=size(Dictionary); lNb=sD2-sD1-1; 
 3. a=Dictionary(:,s);
 4. I=find(a(lNb+2:sD1)>0)+(lNb+1); 
 5. if isempty(I) r=0; return; end  
 6. D=Dictionary(:,[sD2, 1:sD1]);  
 7. Dtilde=diag(1./a(I))*D(I,:);
 8. i=lmv(Dtilde); t=I(i); 
 9. r=Basis(t); 
---------

function nextx=nextset(x)
%input: a {0,1}-n-tuple x i.e. characteristic vector of subset of 1...n
%output:  lexicographically next n-tuple with same number of ones as x; 
%         x itself if x is lexicographically last element. 
 
1. lengthx=length(x);
2. last0=max(find(x==0));  xinit=x(1:last0-1); xfin1s=x(last0+1:lengthx);
3. j=max(find(xinit==1));
4. if isempty(j) nextx=x; return; end
5. nextx=x; nextx([j,j+1])=[0 1];  nextx=[nextx(1:j+1),xfin1s];
6. nextx=[nextx, zeros(1,lengthx-length(nextx))];
---------

function pivot(r,s)
% pivot(r,s) given r in Basis and s in Nbasis, pivots Dictionary for Basis to          
% that for new basis  Basis-r+s according to [A, p183c-1]; updates Basis and Nbasis

 1. global Dictionary Basis Nbasis
 2. sD1=size(Dictionary,1);
 3. a=Dictionary(:,s);  
 4. t=find(Basis==r); tn=find(Nbasis==s); 
 5. rowt=Dictionary(t,:)/a(t); Dictionary(t,:)=rowt;
 6. for i=[1:t-1,t+1:sD1]
 7.  Dictionary(i,:)=Dictionary(i,:)-a(i)*rowt;
 8. end                            
 9. Basis(t)=s; Nbasis(tn)=r;

10. Dictionary=Dictionary.*(abs(Dictionary)>100*eps);
11. Dictionary(:,s)=zeros(sD1,1); Dictionary(t,s)=1;   
---------
\end{verbatim}

\pagebreak

\begin{verbatim}
function [brev,u]=reverse(v)
%  u=reverse(v) 
% in: v in Nbasis.  out:  u=lxminrat(v) and brev\in {0,1}. 
%   brev==1: if v represents an edge and the lexicographic pivot rule applied 
%            to B-u+v generates a pivot back to B. 
%   brev==0: in all other cases. 
 
 1. global Dictionary Basis Nbasis
 2. u=lxminrat(v); 
 3. if u==0  brev=0;  return, end

 4. i=find(Basis==u);      
 5. a=Dictionary(:,v); a0=a(1); ai=a(i);
 6. w0=Dictionary(1,:); wi=Dictionary(i,:); 
 7. wbar=w0-(a0*wi/ai);   

 8. J=Nbasis(Nbasis<u);
 9. if isempty(J) wbarg0=1;  else, wbarg0=all(wbar(J)>=-10*eps); end
10. if (w0(v)>0 & wbarg0) brev=1; else, brev=0;  end
---------

function [xk,maxim]=simplex(c,H,b,initv)
% [xk,maxim]=simplex(c,H,b,initv)
% in: d-row c,  mxd-matrix H, m-column b, dx2-column initv=initvert(H,b), 
%   so that P={x: Hx<=b} is a polytope  having initv(:,1) as a vertex. 
% out: vertex xk of P and real maxim=c*xk=max{cx: Hx<=b} or a message 
%   informing about the unboundedness of the problem. 

 1.    global Dictionary Basis Nbasis
 2.    [Dictionary, Basis, Nbasis]=tratodic(H,b,initv);
 3.    [sD1,sD2]=size(Dictionary);  lNb=sD2-sD1-1;
 4.    u=c*Dictionary(2:lNb+1,sD1+1:sD2);
 5.    Dictionary(1,:)=[1 zeros(1,sD1-1), u(1:lNb), u(sD2-sD1)]; 
    
 6.    [r,j]=slctpivo; 

 7.   while ~isempty(j) 
 8.      if r==0  disp('The problem is unbounded'); return; end 
 9.      pivot(r,Nbasis(j)); 
10.      [r,j]=slctpivo; 
11.   end
      
12. xk=Dictionary(2:lNb+1,sD2)
13.   maxim=Dictionary(1, sD2); 
----------
\end{verbatim}

\pagebreak     

\begin{verbatim}
function [r,j]=slctpivo
% [r,j]=slctpivo: if measuring a non-optimal Dictionary, positive indices  
% r in Basis and j are returned so that r,s=Nbasis(j) reflect the lex pivot 
% selection. A Dictionary is optimal iff j=[] is returned. 

1. global Dictionary Basis Nbasis
2. s=min(find(Dictionary(1,:)<0)); 
3. j=find(Nbasis==s);  
4. if isempty(j) r=0;  return, end
5. r=lxminrat(s); 
---------
function [Dictionary, Basis,Nbasis]=tratodic(H,b,initv)
%[Dictionary, Basis,Nbasis]=tratodic(H,b,initv).  
%in: real mxd-matrix H, m-column b, and dx2-matrix initv=[v,I], with I 
% a    column, of d distinct positive integers so that Hx<=b defines a nonempty 
% polyhedron having v  as a vertex, H(I,:) has rank d, and H(I,:)*v=b(I).
% If existing, initv can be produced by initvert.m 
%out: a dictionary digestible by lrs-function in lrs.m 
 
 1. [m,d]=size(H);  
 2. I=zeros(m,1); I(initv(:,2))=ones(d,1); 
    H=[H(1-I,:); H(I,:)];  b=[b(1-I,:); b(I)]; 
 3. Hupp=H(1:m-d,:); bupp=b(1:m-d);  Hlow=H(m-d+1:m,:); blow=b(m-d+1:m); 
 4. invHlow=inv(Hlow);
 5. Dictionary=[  eye(d)   zeros(d,m-d)      invHlow         invHlow*blow 
                zeros(m-d,d) eye(m-d)   -Hupp*invHlow   bupp-Hupp*invHlow*blow];      
 6. Dictionary=[  1          zeros(1,m) ones(1,d) 0
                 zeros(m,1)       Dictionary        ];
 7. Dictionary=Dictionary.*(abs(Dictionary)>100*eps);
 8.      Basis=1:(m+1); Nbasis=m+2:m+d+1;                  
-----
function [H,b]=vrtoh(V,R)
% [H,b]=vrtoh(V,R) in: vertices as columns of V, ray(directions) as columns in R
%  out: IF vertices define a full dimensional body then a H-description Hx<=b 
%       of the polyhedron. Otherwise an error maessage. 
 
 1. E=V-V(:,1)*ones(1,size(V,2));   E(:,1)=[];  E=[E,R];  
 2. if rank(E)<size(E,1) error('VR-representation is not full dimensional'); end
    
 5. V=-V'; m=size(V,1);   V=[-ones(m,1),V]; 
 6. R=-R'; m=size(R,1);   R=[zeros(m,1),R]; 
 7. A=[V;R]; b=zeros(size(A,1),1);
 8. initv=initvert(A,b);  
 9. [Dictionary, Basis, Nbasis]=tratodic(A,b,initv); 
10. [V,R]=lrs(Dictionary, Basis, Nbasis); 
  
11. Rays=R(:,2:2:size(R,2)); 
12. Rays=Rays'; H=-Rays(:,2:size(Rays,2)); b=Rays(:,1);
\end{verbatim}
{\bf References}: at end of section 3. 

\end{document}